\documentclass[reqno]{amsart}
\usepackage{float}
\newcommand\nc{\newcommand*}  \nc\longnc{\newcommand}
\longnc\VOMIT[1]{#1}          \longnc\OMIT[1]{}
\usepackage[utf8]{inputenc}
\nc\re[1]{(\ref{#1})}   

\nc\lhs{l.h.s. }  \nc\rhs{r.h.s. }
\nc\m[1]{$ #1 $}
\nc\eq[2]{\begin{align} \label{#1} #2 \end{align}}
\nc\eqn[2]{\begin{align} \label{#1} \tagg #2 \end{align}}
\nc\lel[1]{\\ \label{#1}}  
\nc\tagg{\tag*{}}  
\nc\leln[1]{\\ \label{#1} \tagg }  
\nc\mat[1]{\begin{matrix} #1 \end{matrix}}
\nc\smat[1]{\begin{smallmatrix} #1 \end{smallmatrix}}

\nc\0[2]{\ifcase#1{#2}\or\lt(#2\rt)\or\lt[{#2}\rt]\or\lt\{{#2}\rt\}\or
  \mathord<{#2}\mathord>\or\lt\langle{#2}\rt\rangle\or\lt\lvert{#2}\rt
  \rvert\or\lt\lVert{#2}\rt\rVert\fi}
\nc\1[2]{\ifcase#1{#2}\or(#2)\or[#2]\or\{#2\}\or\mathord<{#2}\mathord
  >\or\langle{#2}\rangle\or\lvert{#2}\rvert\or\lVert{#2}\rVert\fi}
\nc\2[2]{\ifcase#1{#2}\or\big(#2\big)\or\big[#2\big]\or\big
  \{#2\big\}\or\big<#2\big>\or\big\langle#2\big\rangle\or\big
  \lvert#2\big\rvert\or\big\lVert#2\big\rVert\fi}
\nc\3[2]{\ifcase#1{#2}\or\Big(#2\Big)\or\Big[#2\Big]\or\Big\{#2\Big
  \}\or\Big<#2\Big>\or\Big\langle#2\Big\rangle\or\Big\lvert#2\Big
  \rvert\or\Big\lVert#2\Big\rVert\fi}
\nc\4[2]{\ifcase#1{#2}\or\bigg(#2\bigg)\or\bigg[#2\bigg]\or\bigg
  \{#2\bigg\}\or\bigg<#2\bigg>\or\bigg\langle#2\bigg\rangle\or\bigg
  \lvert#2\bigg\rvert\or\bigg\lVert#2\bigg\rVert\fi}
\nc\5[2]{\ifcase#1{#2}\or\Bigg(#2\Bigg)\or\Bigg[#2\Bigg]\or\Bigg
  \{#2\Bigg\}\or\Bigg<#2\Bigg>\or\Bigg\langle#2\Bigg\rangle\or\Bigg  
  \lvert#2\Bigg\rvert\or\Bigg\lVert#2\Bigg\rVert\fi}
\nc\9[2]{\ifcase#1{#2}\or\left(#2\right)\or\left[#2\right]\or\left
  \{#2\right\}\or\left\langle{#2}\right\rangle\or\left\langle{#2}\right
  \rangle\or\left\lvert{#2}\right\rvert\or\left\lVert{#2}\right\rVert\fi}
\nc\lt{\mathopen{}\mathclose\bgroup\left} \nc\rt{\aftergroup\egroup\right}

\nc\bi\relax   
\nc\bit{     \kern0.055em}  \nc\biT{     \kern-0.055em}
\nc\bitt{    \kern0.110em}  \nc\biTT{    \kern-0.110em}
\nc\bittt{   \kern0.165em}  \nc\biTTT{   \kern-0.165em}  
\nc\bitttt{  \kern0.220em}  \nc\biTTTT{  \kern-0.220em}  
\nc\bittttt{ \kern0.275em}  \nc\biTTTTT{ \kern-0.275em}  
\nc\bitttttt{\kern0.330em}  \nc\biTTTTTT{\kern-0.330em}

\nc\f\frac                        
\nc\tr{\operatorname{tr}\biT}

\nc\e{\mathrm{e}}  
\nc\dd{\mathop{}\!\mathrm{d}}
  \nc\ddd{\bit\d}  \nc\pd\partial
\def\lta#1{{\overset{{\scriptscriptstyle \leftarrow}}{#1}}}
\def\rta#1{{\overset{{\scriptscriptstyle \rightarrow}}{#1}}}
\def\nablal{\lta{\nabla}}  \def\nablar{\rta{\nabla}}
\nc\tensor\mathbf
\nc\Tensor\boldsymbol
\nc\symm{^{\text{sym}}}
\nc\dev{^{\text{d}}}         \nc\sph{^{\text{s}}}
\nc\qel[1]{#1_{\rm el}}
\nc\para{_{\kern.05em\rule{.03em}{.9ex}\kern.06em\rule{.03em}{.9ex}}}
\nc\ident{\tensor{1}}  \nc\zero{\tensor{0}}
\nc\qcdot{\biT\cdot\biT}
\nc\qdot{^{\biTT\hbox{$\cdot$}}}
\nc\qddot{^{\biTT\hbox{$\cdot\cdot$}}}

\nc\Lapl{\triangle}
\nc\qunderline[1]{\underline{#1\rule[-.7ex]{0em}{0em}}}

\nc\qa{a}           \nc\qan{a_1}             \nc\qann{a_2}
\nc\qc{c}
\nc\qcp{c}
\nc\qe{e}
\nc\qj{q}           \nc\qje{\qj}         \nc\qqje{\tensor{\qj}}
\nc\ql{\ell}
\nc\qr{\tensor{r}}
\nc\qv{v}           \nc\qqv{\mathbf{\qv}}

\nc\qD{D}           \nc\qqD{\tensor{\qD}}
\nc\qE{E}
\nc\qEd{\qE\dev}    \nc\qEs{\qE\sph}
\nc\qL{L}           \nc\qqL{\mathbf{\qL}}
\nc\qT{T}           \nc\qTn{\qT_0}

\nc\qalp{\alpha}
\nc\qgam{\gamma}    \nc\qgamn{\qgam_1^{}}    \nc\qgamnn{\qgam_2^{}}
\nc\qrho{\varrho}
\nc\qsig{\sigma}    \nc\qqsig{\Tensor{\qsig}}

\nc\qlam{k}

\usepackage{makecell}
\usepackage{amsmath}
\usepackage{graphicx}

\begin{document}

\title{Emergence of non-Fourier hierarchies}

\author{Tamás Fülöp $^{1,3}$, Róbert Kovács $^{1,2,3}$, Ádám Lovas $^{1}$, Ágnes Rieth $^{1}$, Tamás Fodor $^{1}$,  Mátyás Szücs $^{1,3}$, Péter Ván $^{1,2,3}$ and Gyula Gróf $^{1}$}

\address{
$^{1}$ \quad Department of Energy Engineering, Faculty of Mechanical Engineering, BME, Budapest, Hungary\\
$^{2}$  \quad Department of Theoretical Physics, Wigner Research Centre for Physics,
Institute for Particle and Nuclear Physics, Budapest, Hungary \\
$^3$ \quad Montavid Thermodynamic Research Group
}

\begin{abstract}The non-Fourier heat conduction phenomenon on room temperature is analyzed from various aspects. The first one shows its experimental side, in what form it occurs and how we treated it. It is demonstrated that the Guyer-Krumhansl equation can be the next appropriate extension of Fourier's law for room temperature phenomena in modeling of heterogeneous materials. 
The second approach provides an interpretation of generalized heat conduction equations using a simple thermomechanical background. Here, Fourier heat conduction is coupled to elasticity via thermal
expansion, resulting in a particular generalized heat equation for the
temperature field. Both of the aforementioned approaches show the size dependency of non-Fourier heat conduction.
Finally, a third approach is presented, called pseudo-temperature modeling. It is shown that non-Fourier temperature history can be produced by mixing different solutions of Fourier's law. That kind of explanation indicates the interpretation of underlying heat conduction mechanics behind non-Fourier phenomena. 
\end{abstract}
\maketitle

\pagestyle{plain}
\section{Introduction}
The Fourier's law \cite{Fou822} 
\begin{align}
\mathbf q = - k \nablar T
\end{align}
is one of the most applicable, well-known elementary physical laws in engineering practice. Here, $\mathbf q$ is the heat flux vector, $T$ is absolute temperature, $k$ is thermal conductivity.
However, as all the constitutive equations, it also has limits of validation. Phenomena that do not fit into these limits, called non-Fourier heat conduction, appear in many different forms. 
Some of them occur at low temperature like the so-called second sound and ballistic (thermal expansion induced) propagation \cite{Tisza38, JosPre89, JosPre90a, Chen01, VanFul12, KovVan15}. These phenomena have been experimentally measured several times \cite{Acketal66, JacWal71, Pesh44, McN74t} and many generalized heat equations exist to simulate them \cite{DreStr93a, MulRug98, FriCim95, KovVan16, KovVan18, BarSte05a, HerBec00}. The success in low-temperature experiments resulted in the extension of this research field to find the deviation at room temperature as well. One of the most celebrated result is related to Mitra et al. \cite{MitEta95} where the measured temperature history was very similar to a wave-like propagation. However, these results have not been reproduced by anyone and undoubtedly demanded for further investigation. 

In most of the room-temperature measurements, the existence of Maxwell-Cattaneo-Vernotte (MCV) type behavior attempted to be proved \cite{Cattaneo58, Vernotte58}. It is this MCV equation that is used to model the aforementioned second sound, the dissipative wave propagation form of heat \cite{JosPre89, Tisza47, Lan47}. The validity of MCV equation for room temperature behavior has not yet been justified, despite of the numerous experiments.
It is important to note that many other extensions of Fourier equation exist beyond the MCV one, such as the Guyer-Krumhansl (GK) equation \cite{GuyKru66a1, GuyKru66a2, Van01a, Zhu16a, Zhu16b}, the dual phase lag model \cite{Tzou96}, and their modifications, too \cite{KovVan15, SellEtal16, RogEtal17}. Some of these possess stronger physical background, some others not \cite{FabEtal16, Ruk17, KovVan18dpl}. 

The simplest extension of MCV equation is the GK model, which reads:
\begin{align}
\tau \dot {\mathbf q }+ \mathbf q + k \nablar T - \kappa^2 \Lapl \mathbf q =0, \label{GK}
\end{align}
where the coefficient $\tau$ is called relaxation time and $\kappa^2$ is regarded as a dissipation parameter and the dot denotes the time derivative. This GK-type constitutive equation contains the MCV-type by considering $\kappa^2=0$ and the Fourier equation taking $\tau=\kappa^2=0$. This feature of GK equation allows to model both wave-like temperature history and over-diffusive one. This is more apparent when one applies the balance equation of internal energy in order to eliminate $\mathbf q$:
\begin{align}
\rho c \dot T + \nablar \cdot \mathbf q = 0, \label{enbal}
\end{align}
with mass density $\rho$, specific heat $c$ and volumetric source neglected, one obtains
\begin{align}
\tau \ddot T + \dot T = a \Lapl T + \kappa^2 \Lapl \dot T, \label{GKT}
\end{align}
with thermal diffusivity $a =k / (\rho c)$. One can realize that equation (\ref{GKT}) contains the Fourier heat equation 
\begin{align}
\dot T = a \Lapl T 
\label{Fouriereq}
\end{align}
as well as its time derivative, with different coefficients. It becomes more visible after rearranging eq.~(\ref{GKT}):
\begin{align}
\tau \left (\dot T - \frac{\kappa^2}{\tau} \Lapl T \right )^{.} + \dot T - a \Lapl T = 0.
\label{HierGKT}
\end{align}
When the so-called \cite{Botetal16, Vanetal17} Fourier resonance condition $\kappa^2/\tau = a$ holds, the solutions of the Fourier equation (\ref{Fouriereq}) are covered by the solutions of (\ref{GKT}). Meanwhile, when $\kappa^2<a \tau$ the wave-like behavior is recovered and this domain is called as under-damped region. In the opposite case ($\kappa^2>a \tau$), there is no visible wave propagation and it is called over-diffusive (or over-damped) region. We measured the corresponding over-diffusive effect several times in various materials such as metal foams, rocks and in a capacitor, too \cite{Botetal16, Vanetal17}. Furthermore, a similar temperature history has been observed in a biological material \cite{KovVan18dpl}. 

In this paper, further aspects of over-diffusive propagation are discussed. In the following sections the size dependence of the observed over-damped phenomenon is discussed both  experimentally and theoretically. Moreover, the approach of pseudo-temperature is presented in order to provide one concrete possible interpretation for non-Fourier heat conduction. 

\section{Size dependence}

Our measurements reported here are performed on basalt rock samples with three different thicknesses, $1.86$, $2.75$ and $3.84$ mm, respectively. We have applied the same apparatus of heat pulse experiment as described in \cite{Botetal16, Vanetal17}, schematically depicted in Fig.~\ref{expsetup} below. 

\begin{figure}[H]
\centering
\includegraphics[width=8cm]{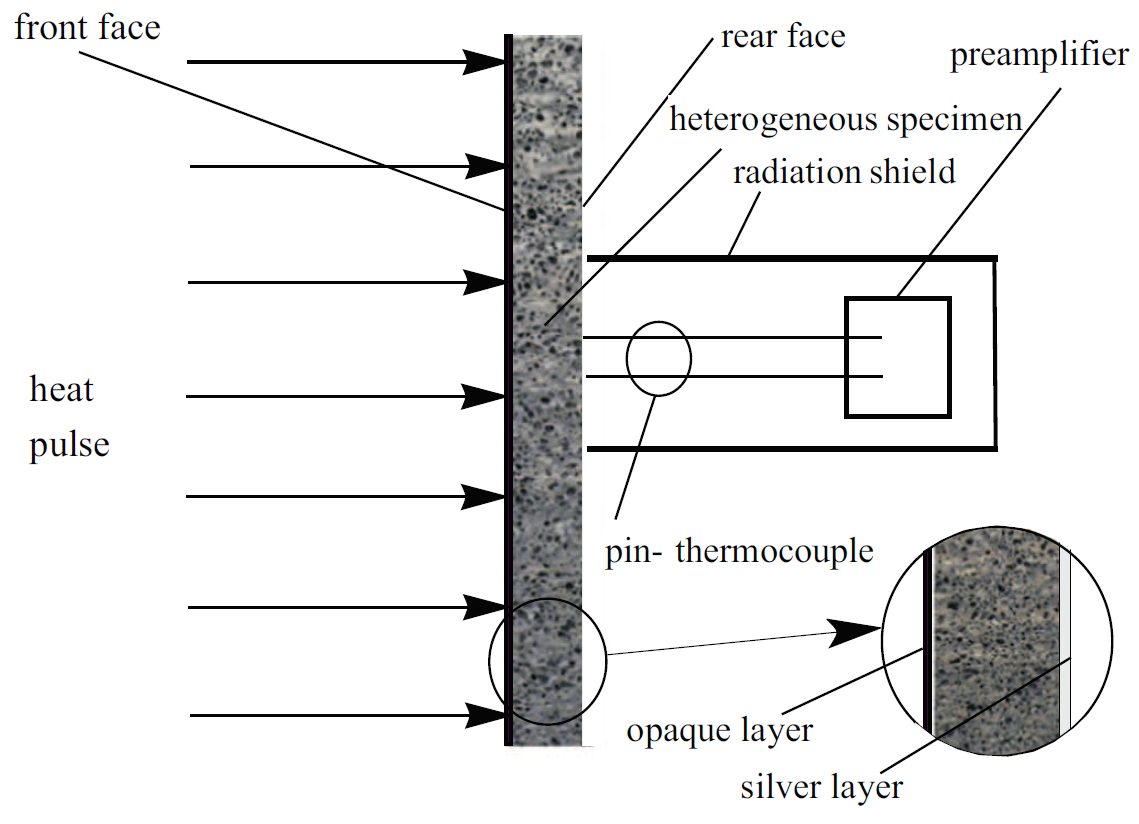}
\caption{Setup of our heat pulse experiment \cite{Vanetal17}. }
\label{expsetup}
\end{figure}   

\begin{figure}[H]
\centering
\includegraphics[width=15cm]{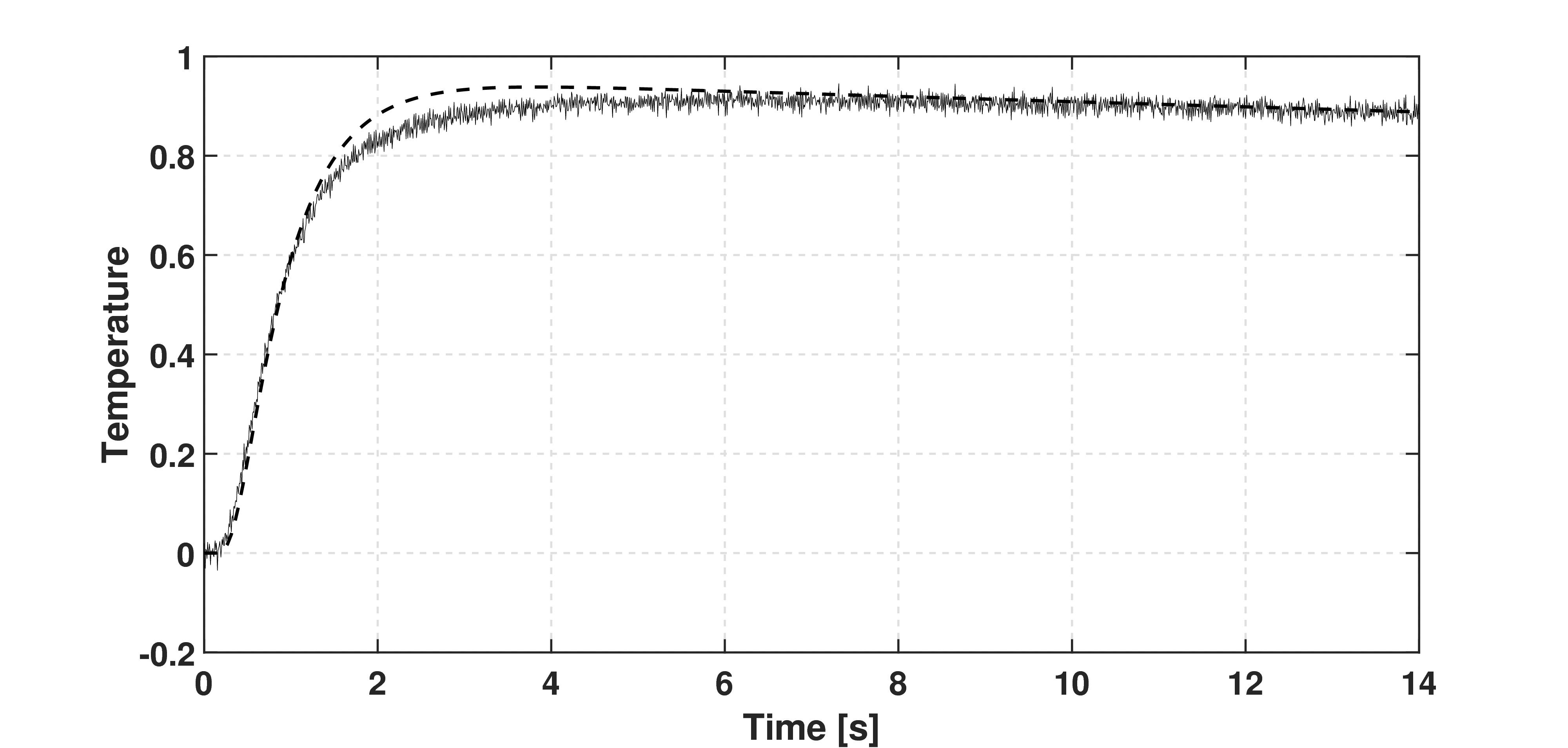}
\caption{Data recorded for basalt rock sample with thickness of $1.86$ mm. The dashed line shows the prediction of Fourier's law.}
\label{expfou1}
\end{figure}

\begin{figure}[H]
\centering
\includegraphics[width=15cm]{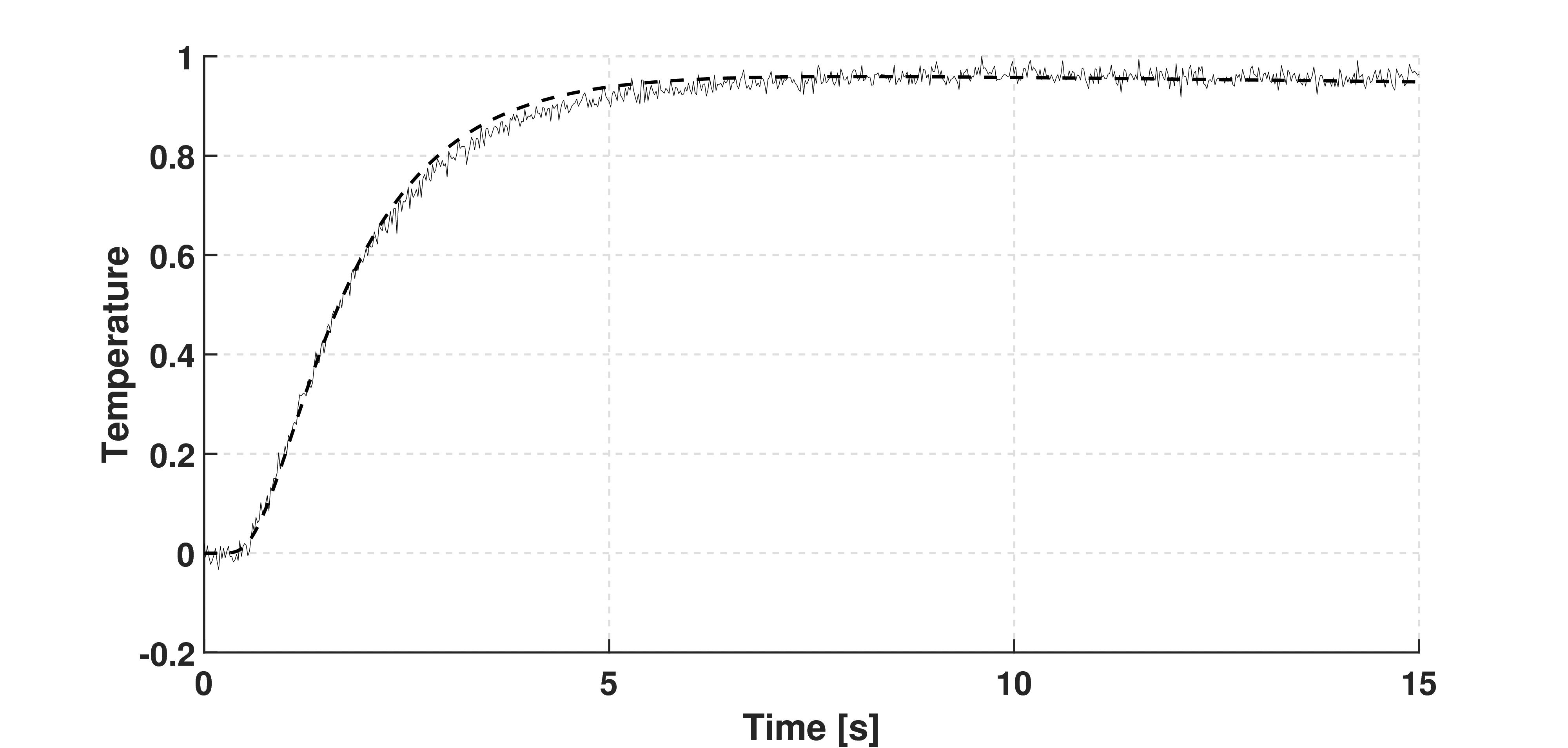}
\caption{Data recorded for basalt rock sample with thickness of $2.75$ mm. The dashed line shows the prediction of Fourier's law.}
\label{expfou2}
\end{figure}

\begin{figure}[H]
\centering
\includegraphics[width=15cm]{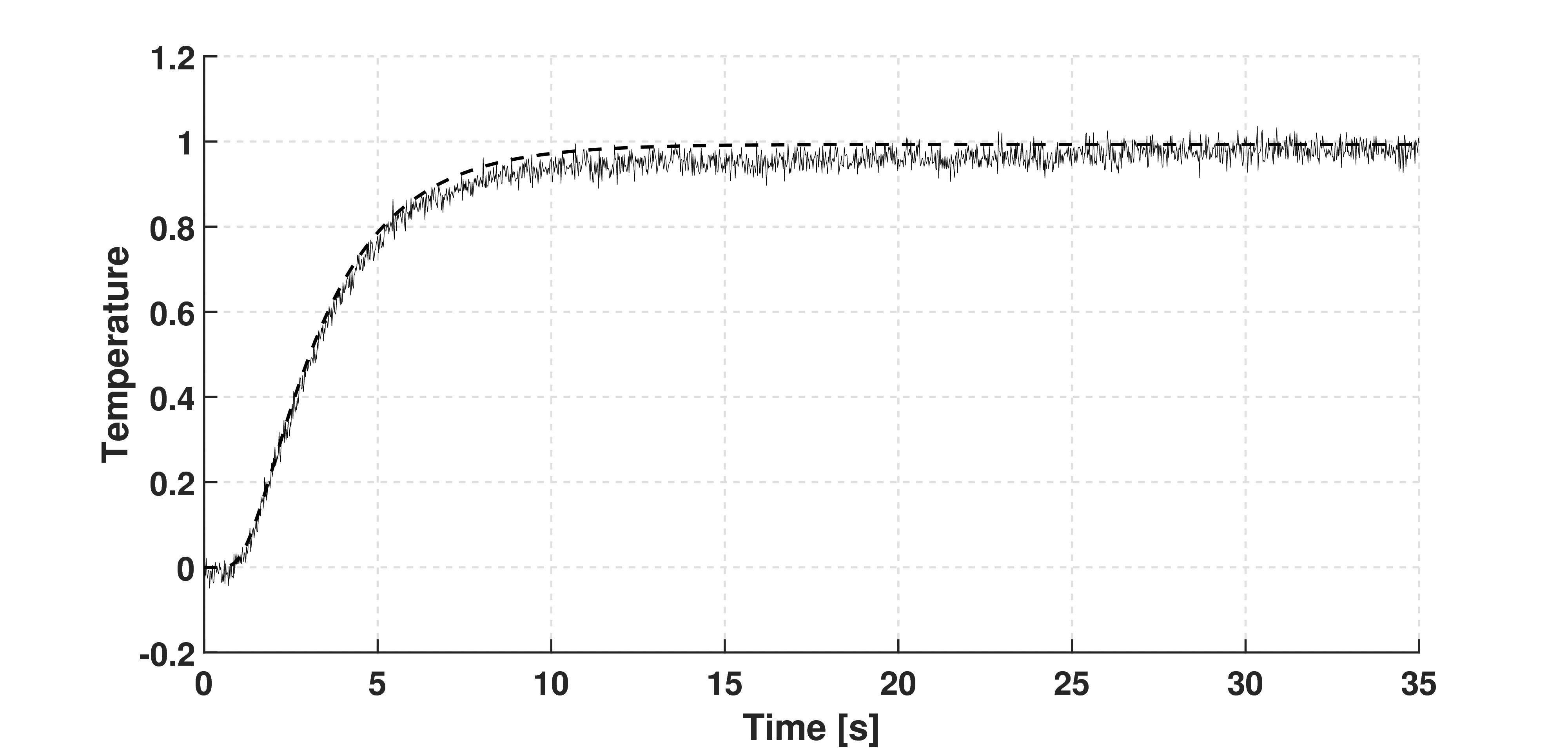}
\caption{Data recorded for basalt rock sample with thickness of $3.84$ mm. The dashed line shows the prediction of Fourier's law.}
\label{expfou3}
\end{figure}

In each case, the rear-side temperature history was measured and numerically evaluated solving the GK equation. The recorded dimensionless temperature signals are plotted in Figs.~\ref{expfou1}, \ref{expfou2}, \ref{expfou3}. In these figures, the dashed line shows the solution of Fourier equation using thermal diffusivity corresponding to the initial part of temperature rising on the rear side. It is clear that the measured signal deviates from the Fourier-predicted one even with considering non-adiabatic (cooling) boundary condition. That deviation weakens with increasing the sample thickness, for the thickest one it is hardly visible and the prediction of Fourier's law is almost acceptable. 

\begin{figure}[H]
\centering
\includegraphics[width=15cm]{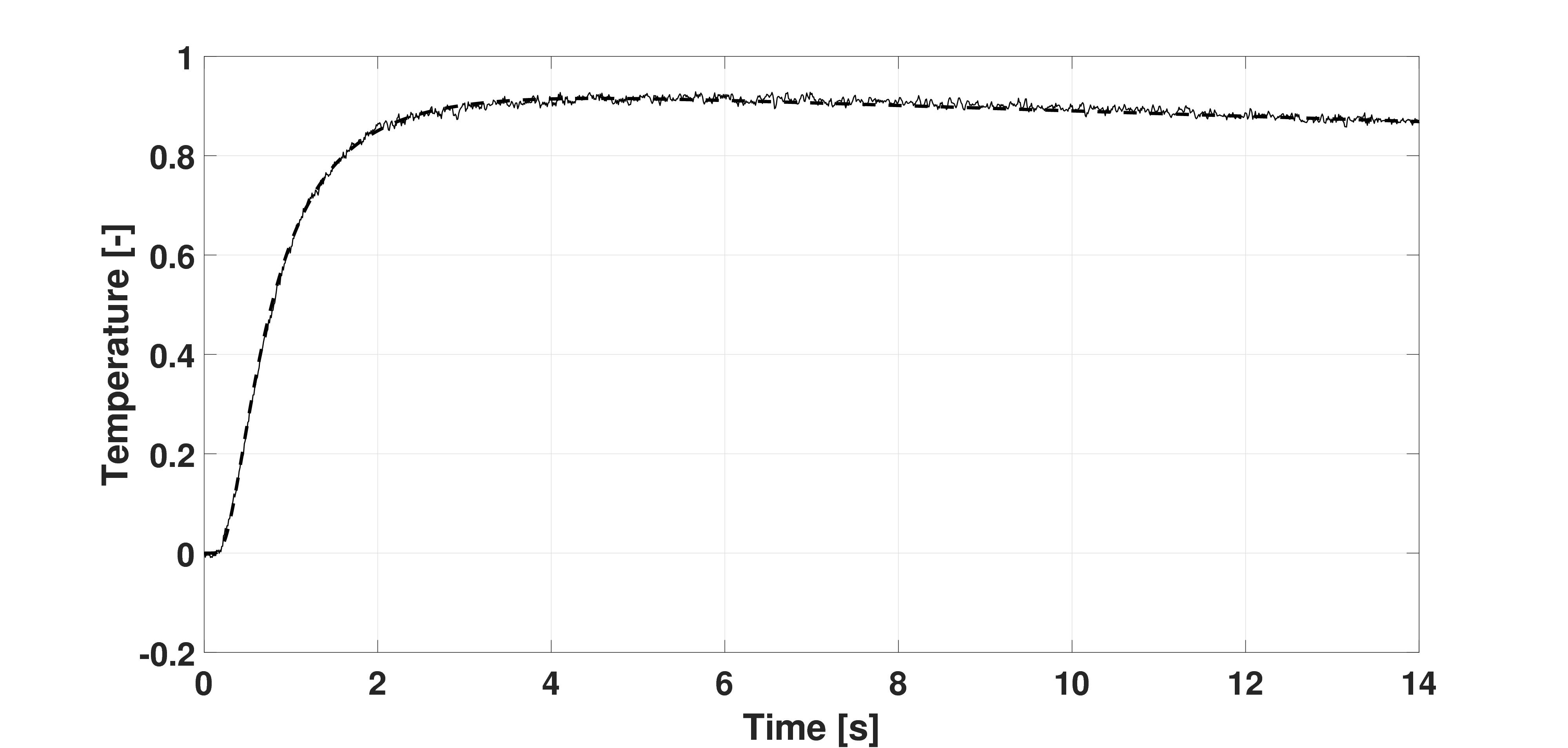}
\caption{Data recorded using the basalt with thickness of $1.86$ mm. The solid continous line shows the prediction of GK equation.}
\label{expfou4}
\end{figure}

The evaluation of the thinnest sample using the Guyer-Krumhansl equation is shown in Fig.~\ref{expfou4}. The fitted coefficients are summarized in Table \ref{expcoeff}. 

\begin{table}[H]
\caption{Summarized results of fitted coefficients in Fourier and GK equations.} \label{expcoeff}
\centering
\begin{tabular}{c|c|ccc}
\textbf{\makecell{Thickness \\ $L$, $[\rm mm]$}} &
\textbf{\makecell{Fourier \\ thermal diffusivity \\ $a_F$,  $\cdot 10^{-6} \big[\frac{\rm m^2}{\rm s}\big ]$ }}	& \textbf{\makecell{Guyer-Krumhansl \\ thermal diffusivity\\ $a_{GK}$,  $\cdot 10^{-6} \big[\frac{\rm m^2}{\rm s}\big ]$}}	& \textbf{\makecell{Relaxation \\ time \\ $\tau$, $[\rm s]$}} & \textbf{\makecell{Dissipation \\ parameter \\ $\kappa^2$, $\cdot 10^{-6} [\rm m^2]$}}\\
$1.86$ &$0.62$		& $0.55$			& $0.738$		& $0.509$ \\
$2.75$&$0.67$			& $0.604$			& $0.955$		& $0.67$\\
$3.84$&$0.685$	& 	$0.68$		& 	$0.664$		& $0.48$\\
\end{tabular}
\end{table}

Deviation from the Fourier prediction is weak but is clearly present, and has size dependent attributes. Concerning the ratio of parameters, i.e., investigating how considerably the Fourier resonance condition $a \tau / \kappa^2 = 1$ is violated, the
outcome can be seen in Table \ref{expcoeff2}. As analysis of the results, it is remarkable to note the deviation of the GK
fitted thermal diffusivity from the Fourier fitted one, and that this deviation is size dependent. For the thickest sample, which can be well described by Fourier's law, the fitted thermal diffusivity values are practically equal, and the ratio of parameters is very close to the Fourier resonance value 1. 

\begin{table}[H]
\caption{Ratio of the fitted coefficients.} \label{expcoeff2}
\centering
\begin{tabular}{cc}
\textbf{\makecell{Thickness \\ $L$, $[\rm mm]$}} &
\textbf{\makecell{Ratio of parameters \\ $\frac{a_{GK} \tau}{\kappa^2} $  }}\\
$1.86$ &	$0.804$	\\
$2.75$&		$0.854$	\\
$3.84$&		$0.943$	\\
\end{tabular}
\end{table}

The next section is devoted to a possible explanation for the emergence of a
generalized heat equation with higher time and space derivatives. All
coefficients of the higher time and space derivative terms are related to
well-known material parameters. The result also features size dependent
non-Fourier deviation.

\section{Seeming non-Fourier heat conduction induced by elasticity
coupled via thermal expansion}

While, in general, one does not have a direct physical interpretation of
the phenomenon that leads to, at the phenomenological level, non-Fourier
heat conduction, here follows a case where we do know this
background phenomenon.
Namely, in case of heat conduction in solids, a plausible possibility
is provided by an interplay between elasticity and thermal expansion.
Namely, without thermal expansion, elasticity -- a tensorial behaviour
-- is not coupled to Fourier heat conduction -- a vectorial one -- in
isotropic materials. However, with nonzero thermal expansion, strains
and displacements have to be in accord both with what elastic mechanics
dictates and with what position dependent temperature imposes. The
coupled set of equations of Fourier heat conduction, of elastic
mechanics and of kinematic relationships, after eliminating the
kinematic and mechanical quantities, leads to an equation for
temperature only that contains higher derivative corrections to
Fourier's equation. It is important to check how remarkable these
corrections are. In the following section we present this derivation and
investigation.

\subsection{The basic equations}  \label{.62..2.2.}

In all respects involved, we choose the simplest assumptions: the small-strain regime, a Hooke-elastic homogeneous and isotropic solid material, with constant thermal expansion coefficient, essentially being at rest
with respect to an inertial reference frame. Kinematic, mechanical and
thermodynamical quantities and their relationships are considered along
the approach detailed in \cite{Godollo-en,MMAS,IWNET}.

The Hooke-elastic homogeneous and isotropic material model states, at
any position \m { \qr }, the constitutive relationship
 \eq{.62.1.}{
\qqsig\dev = \qEd \qqD\dev ,
 \quad
\qqsig\sph = \qEs \qqD\sph ,
 \qquad \quad
\qEd = 2G ,
 \quad
\qEs = 3K ,
 }
 \eq{.62.2.}{
\qqsig = \qEd \qqD\dev + \qEs \qqD\sph = \qEd \qqD + \9 1 {
\qEs - \qEd } \qqD\sph
 }
between stress tensor \m { \qqsig } and elastic deformedness tensor
\m { \qqD } (which, in many cases, coincides with the strain tensor), where \m { {}\dev } and \m{ {}\sph }
denote the deviatoric (traceless) and spherical (proportional to the
unit tensor \m { \ident }) parts, i.e.,
 \eq{.62.3.}{
\qqD\sph = \f {1}{3}\9 1 { \tr \qqD } \ident ,
 \quad
\qqD\dev = \qqD - \qqD\sph ;
\qquad  \text{hence, e.g.,}  \quad
\ident\sph = \ident ,
 \quad
\ident\dev = \zero .
 }
Stress induces a time derivative in the velocity field \m { \qqv } of
the solid medium, according to the equation
 \eq{.62.4.}{
\qrho \dot \qqv = \qqsig \qcdot \nablal
 }
with mass density \m { \qrho } being constant in the small-strain
regime. For the velocity gradient \m { \qqL } and its symmetric part,
one has
 \eq{.62.5.}{
\qqL = \qqv \otimes \nablal ,
 \quad
\tr \qqL\symm = \tr \qqL = \qqv \qcdot \nablal ,
 \quad
\0 1 { \qqL\symm }\sph = \f {1}{3} \9 1 { \tr \qqL\symm } \ident = \f
{1}{3} \9 1 { \qqv \qcdot \nablal } \ident ,
 }
 \eq{.62.6.}{
\0 1 { \qqL\symm \qcdot \nablal } \qcdot \nablal & = \f {1}{2}
\pd_i \pd_j \0 1 { \pd_i \qv_j + \pd_j \qv_i } = \f {1}{2} \0 2 { \Lapl
\0 1 { \nablar \qcdot \qqv } + \Lapl \0 1 { \nablar \qcdot \qqv
} } = \Lapl \0 1 { \qqv \qcdot \nablal } ,
 \lel{.62.7.}
 \rule{0em}{4.ex}
\0 1 { \qqL \qcdot \nablal } \qcdot \nablal & =  \Lapl \0 1
{ \qqv \qcdot \nablal } ,
 }
where the Einstein summation convention for indices has also been
applied. Again using this convention, and the Kronecker delta notation,
to any scalar field \m { f },
 \eq{.63.8.}{
\pd_j \9 1 { f \delta_{ij} } = \delta_{ij} \pd_j f = \pd_i f ,
\qquad
\9 1 { f \ident } \qcdot \nablal = \nablar f
 }
follow, which are also to be utilized below.

The small-deformedness relationship among the kinematic quantities, with
linear thermal expansion coefficient \m { \qalp } considered constant,
and absolute temperature \m { \qT }, is
 \eq{.63.9.}{
\qqL\symm = \dot\qqD + \qalp \dot\qT \ident \,.
 }
For specific internal energy \m { \qe },
 \eq{@9773}{
\qe = \qc \qT + \f {\qEs \qalp}{\qrho} \qT \tr \qqD\sph + \qel\qe \, ,
 \quad
\qel\qe = \f {\qEd}{2 \qrho} \tr \9 2 { \9 1 { \qqD\dev}^2 } +
 \f {\qEs}{2 \qrho} \tr \3 2 { \9 1 { \qqD\sph}^2 } \, ,
 }
its balance,
 \eq{@10018}{
\qrho \dot\qe = \tr \9 1 { \qqsig \qqL } - \qqje \qcdot \nablal \, ,
 }
after subtracting the contribution \m { \qrho \qel{\dot{\qe}} } coming
from specific elastic energy \m { \qel{\qe} } and the corresponding
elastic part \m{ \tr \2 1 { \qqsig \dot \qqD } } of the mechanical power
\m { \tr \9 1 { \qqsig \qqL } }, is
 \eq{.63.10.}{
\qrho \9 1 { \qe - \qel{\qe} }\qdot = \qrho \qc \dot\qT + \qEs \qalp
\qTn \tr \dot \qqD\sph = \mathop- \qqje \qcdot \nablal \,,
 \qquad
\mbox{with}
 \quad
\qqje = - \qlam \nablar \qT \,,
 }
where \m { \qc } is specific heat corresponding to constant zero stress
(or pressure), temperature has been approximated in one term of \re{.63.10.} by an
initial homogeneous absolute temperature value \m { \qTn } to stay in accord with
the linear (small-strain) approximation, and heat flux \m { \qqje }
follows the Fourier heat conduction constitutive relationship with
thermal conductivity \m { \qlam } also treated as a constant.

\subsubsection{The derivation}  \label{.63..2.3.}

The strategy is to eliminate \m { \qqsig } in favour of (with the aid
of) \m { \qqD }, then \m { \qqD } is eliminated in favour of \m {
\qqL\symm }, after which we can realize that both from the mechanical
direction and from the thermal one we obtain relationship between \m {
\qqv \qcdot \nablal } and \m { \qT }, which, eliminating \m { \qqv
\qcdot \nablal }, yields an equation for \m { \qT } only.

Starting with the thermal side,
 \eq{@11.}{
\qrho \qc \dot\qT + \qEs \qalp \qTn \tr \2 1 { \qqL\symm - \qalp \dot\qT
\ident }\sph & = \qrho \qc \dot\qT + \qEs \qalp \qTn \0 1 { \qqv \qcdot
\nablal } - \qEs \qalp^2 \qTn \dot\qT \cdot 3 =
 \tagg\lel{@12.}
& = \2 1 { \underbrace{ \qrho \qc - 3 \qEs \qalp^2 \qTn }_{\qgamn} }
\dot\qT + \qEs \qalp \qTn \1 1 { \qqv \qcdot \nablal }
 \,,
 \leln{@13.}
& = {} - \qqje \qcdot \nablal  = - \0 1 { - \qlam  \nablar \qT }
\qcdot \nablal = \qlam \Lapl \qT
 \quad \Longrightarrow
 \lel{.63.11.}
\qEs \qalp \qTn \0 1 { \qqv \qcdot \nablal } & = \qlam \Lapl \qT -
\qgamn \dot\qT \,.
 }
Meanwhile, from the mechanical direction, aiming at being in tune with
\re{.63.11.}:
 \eqn{@15.}{
\qEs \qalp \qTn \0 1 { \ddot\qqv \cdot \nablal } & = \qEs \qalp \qTn
\f {1}{\qrho} \0 1 { \dot\qqsig \cdot \nablal } \cdot \nablal =
 \leln{@16.}
& = \f {\qEs \qalp \qTn}{\qrho} \9 3 { \9 2 { \qEd \dot\qqD + \9 1 {
\qEs - \qEd } \dot\qqD\sph } \cdot \nablal } \cdot \nablal =
 \leln{@17.}
& = \f {\qEs \qalp \qTn}{\qrho}
\biggl\{
\biggl[
{ \qEd \9 1 { \qqL\symm - \qalp \dot\qT \ident } + }
 \leln{@18.}
& \hskip 6.3 em + \9 1 { \qEs - \qEd } \9 1 { \qqL\symm - \qalp \dot\qT
\ident }\sph
\biggr]
\cdot \nablal
\biggr\}
\cdot \nablal =
 \leln{@19.}
& = \f {\qEs \qalp \qTn}{\qrho}
\biggl\{
\biggl[
\qEd \qqL\symm - \qEd \qalp \dot\qT \ident + \9 1 { \qEs - \qEd }
\f{1}{3} \0 1 { \qqv \qcdot \nablal } \ident \mathrel-
 \leln{@20.}
 & \hskip 6.3 em
- \9 1 { \qEs - \qEd } \qalp \dot\qT \ident
\biggr]
\cdot \nablal
\biggr\}
\cdot \nablal =
 \leln{@21.}
& = \f {\qEs \qalp \qTn}{\qrho} \9 2 { \qEd \Lapl \0 1 { \qqv
\qcdot \nablal } + \f{ \qEs - \qEd }{3} \Lapl \0 1 { \qqv
\qcdot \nablal } - \qEs \qalp \Lapl \dot\qT } =
 \leln{@22.}
& = \f {\qEs \qalp \qTn}{\qrho} \9 2 { \f{ \qEs + 2 \qEd }{3} \Lapl \0 1
{ \qqv \qcdot \nablal } - \qEs \qalp \Lapl \dot\qT } =
 \leln{@23.}
& = \f{ \qEs + 2 \qEd }{3\qrho} \Lapl \9 2 { \qEs \qalp \qTn \0 1 {
\qqv \qcdot \nablal } } - \f{ \1 1 {\qEs \qalp}^2 \qTn }{\qrho} \Lapl
\dot\qT =
 \leln{@24.}
& = \underbrace{ \f{ \qEs + 2 \qEd }{3\qrho} }_{\qcp\para^2} \Lapl \9 1
{ \qlam \Lapl \qT - \qgamn \dot\qT } - \f{ \1 1 {\qEs \qalp}^2 \qTn
}{\qrho} \Lapl \dot\qT \,;
 \quad \text{in parallel,}
 \lel{.64.12.}
& = \9 1 { \qlam \Lapl \qT - \qgamn \dot\qT }\qddot = \qlam \Lapl \ddot
\qT - \qgamn \dddot\qT
 \qquad \text{\1 2 {cf.\ \re{.63.11.}}}
 }
(where \m { \qcp\para } is the longitudinal elastic wave propagation
velocity);
hence, summarizing the final result in two equivalent forms,
 \eq{.64.13.}{
\9 1 { \qgamn \dot\qT - \qlam \Lapl \qT }\qddot & = \qcp\para^2 \Lapl \9
1 { \qunderline{\qgamn} \dot\qT - \qlam \Lapl \qT } + \f{ \1 1 {\qEs
\qalp}^2 \qTn }{\qrho} \Lapl \dot\qT \,,
 \lel{.64.14.}
\qgamn \9 1 { \ddot\qT - \qunderline{ \qcp\para^2 } \Lapl \qT }\qdot & =
\qlam \Lapl \qgamn \9 1 { \ddot\qT - \qcp\para^2 \Lapl \qT } + \f{ \1 1
{\qEs \qalp}^2 \qTn }{\qrho} \Lapl \dot\qT \,.
 }
The first form here tells us that we have here the wave equation of a
heat conduction equation, the last term on the \rhs somewhat detuning
the heat conduction equation of the \rhs with respect to the one on
the l.h.s. (the underlined coefficient is the one becoming modified when
its term is melted together with the last term). In the meantime, the
second form shows the heat conduction equation of a wave equation, the
last term on the \rhs detuning the underlined coefficient.

Both forms show that coupling, after elimination, leads to a hierarchy
of equations, with an amount of detuning that is induced by the coupling
-- for similar further examples, see \cite{hierar}.

We close this section by rewriting the final result in a form
that enables to estimate the contribution of thermal expansion coupled
elasticity to heat conduction:
 \eq{@16989}{
\f{1}{\qcp\para^2} \9 1 { \qgamn \dot\qT - \qlam \Lapl \qT }\qddot & =
\Lapl \9 2 { \9 1 { {\qgamn} + \f{ \1 1 {\qEs \qalp}^2 \qTn }{\qrho
\qcp\para^2} } \dot\qT - \qlam \Lapl \qT } \,,
 }
i.e.,
 \eq{@17339}{
\f{1}{\qcp\para^2} \9 1 { \qgamn \dot\qT - \qlam \Lapl \qT }\qddot & =
\Lapl \4 2{ \4 1 { \underbrace{ \qrho \qc - \f{ 6 \qEd \qEs \qalp^2 \qTn
}{\qEs + 2 \qEd} }_{\qgamnn} } \dot\qT - \qlam \Lapl \qT } \,.
 }
One message here is that, thermal expansion coupled elasticity modifies
the thermal diffusivity \m { \qa = \qlam / \1 1 {\qrho \qc} } to an effective
one \m { \qann = \qlam / \qgamnn = \1 1 {\qrho \qc / \qgamnn} \cdot
\qa } (see the heat conduction on the \rhs). For metals, this means a
few-percent shift (1\% for steel and copper, and 6\% for
aluminum) at room temperature.

The other is that, for a length scale (e.g., characteristic sample size)
\m { \ql } and the corresponding Fourier time scale \m { \ql^2/\qa },
the \rhs\ is, to a (very) rough estimate, \m { 1/\ql^2 } times a heat
conduction equation while the \lhs\ is (similarly roughly)
 \eq{@18137}{
\f{1}{ \1 1 {\ql^2/\qa}^2 } \cdot \f{1}{\qcp\para^2}
 }
times the (nearly) same heat conduction equation (a one with \m {
\qan = \qlam / \qgamn }). In other words, the \lhs\ provides a
contribution to the \rhs\ via a dimensionless factor
 \eq{@18435}{
\f{\ql^2}{ \1 1 {\ql^2/\qa}^2 } \cdot \f{1}{\qcp\para^2} = 
\f {\qa^2}{\ql^2 \qcp\para^2} .
 }
This dimensionless factor is about \m { 10^{-10} } to \m{ 10^{-13} } for
metals, \m { 10^{-14} } for rocks and \m { 10^{-15} } for plastics with
\m { \ql = 3 \bittt \text{mm} }, a typical size for flash experiments.
Therefore, the effect of the \lhs\ appears to be negligible with respect
to the \rhs.

It is important to point out that the first phenomenon---the emergence
of effective thermal diffusivity---would remain unnoticed in the
analogous
one space dimensional calculation:
 \eq{@16883}{
\qsig = \qE \qD ,
 \qquad
\qrho \dot \qv & = \qsig' ,
 \qquad
\qL = \qv' = \dot \qD + \qalp \dot \qT ,
 \lel{@17043}
\qje = - \qlam \qT' ,
 \qquad
\qe & = \qc \qT + \f {\qE \qalp}{\qrho} \qT \qD + \f {\qE}{2 \qrho} \qD^2
 \qquad
\Longrightarrow
 \lel{@17174}
\f {\qrho}{\qE} \9 2 { \2 1 { \qrho \qc - \qE \qalp^2 \qTn } \dot\qT -
\qlam \qT'' } \qddot & =
\9 2 { \qrho \qc \dot\qT - \qlam \qT'' }''
 }
[no detuning of \m { \qrho \qc } on the \rhs].
It is revealed only in the full 3D treatment, which enlights possible
pitfalls of 1D considerations in general as well.

As conclusion of this section, thermal expansion coupled elasticity may introduce a
few percent effect (a material dependent but sample size independent value) in determining
thermal diffusivity from flash experiments or other transient processes
(while its other consequences may be negligible).

\section{Pseudo-temperature approach}

The experimental results serve to check whether a certain theory used for describing the observed phenomenon is acceptable or not. The heat pulse (flash) experiment results may show various temperature histories. Generally the flash measurement results are according to the Fourier theory. In some cases, as reported in \cite{Botetal16, Vanetal17} the temperature histories show “irregular” characteristics, especially these histories could be described by the help of various non-Fourier models \cite{JouEtal15, SellEtal16, JouCimm16, KovVan15}. Some kind of non-Fourier behaviour could be constructed as it is shown in the following. This is only an illustration how two parallel Fourier mechanisms could result a non-Fourier-like temperature history. The idea is strongly motivated by the hierarchy of Fourier equations in the GK model \cite{VanKovFul15} as mentioned previously, however, their interaction is not described in detail. 

The sample that we investigate now is only a hypothetic one, we may call it as a “pseudo-matter”. We consider in the following that the pseudo-matter formed by parallel material strips is wide enough that the interface effects might be neglected, i.e., they are like insulated parallel channels. We also consider that only the thermal conductivities are different, and the strips have the same mass density and specific heat. During the flash experiment after the front side energy input, a simple temperature equalisation process happens in the sample in case of adiabatic boundary conditions. Since the flash method is widely developed, the effects of the real measurement conditions (heat losses, heat gain, finite pulse time, etc.) are well treated in the literature.

\begin{figure}[H]
\centering
\includegraphics[width=15cm]{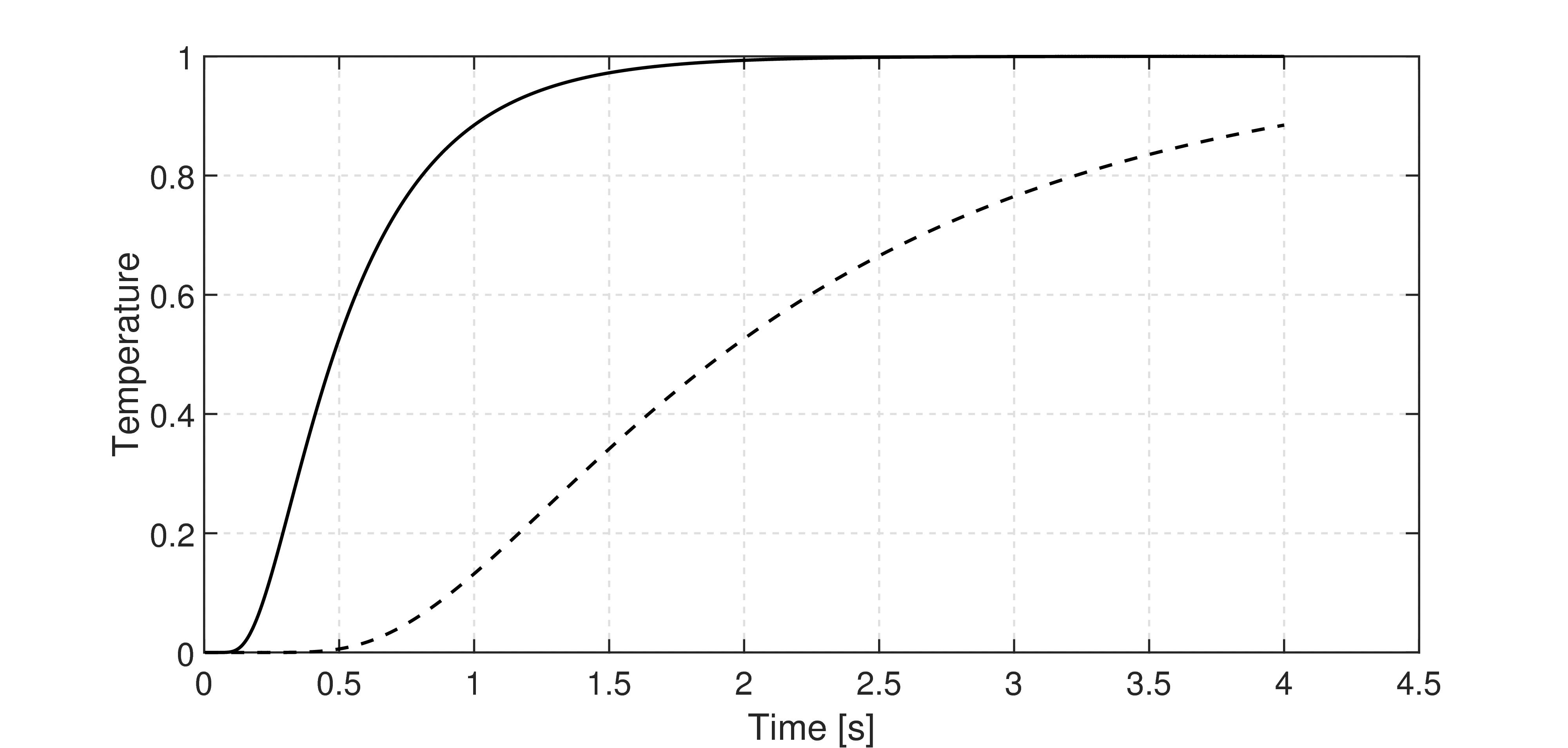}
\caption{Rear-side temperature history; solid line: $a=10^{-6} \ \text{m$^2$/s}$, dashed line:  $a=2.5 \cdot 10^{-7} \ \text{m$^2$/s}$, $L=2 \ \text{mm}$.}
\label{pseudo1}
\end{figure}

Figure \ref{pseudo1} shows two temperature histories with thermal diffusivities of different magnitude, both of them are the solution of Fourier heat equation.
The mathematical formula that expresses the temperature history of the rear
side in the adiabatic case is \cite{ParEtal61}:
\begin{align}
\nu(\xi=1, Fo)=1 + 2 \sum\limits^{\infty}_{m=1} (-1)^m e^{-(m^2 \pi^2 Fo)},
\end{align}
where $\nu$ is the dimensionless temperature, $\xi$ is the normalized spatial coordinate ($\xi=1$ corresponds to the rear-side) and $Fo = a \cdot t /(L^2)$ stands for the Fourier number (dimensionless time variable). This is an infinite series with property of slow convergence for short initial time intervals. An alternative formula derived using the
Laplace theorem to obtain faster convergence for $Fo < 1$ \cite{James80}:
\begin{align}
p(Fo) =\frac{2}{\sqrt{\pi Fo}}\sum \limits^{\infty}_{n=0} e^{-\frac{(2n+1)^2}{4 Fo}}.
\label{pfo}
\end{align}
In the further analysis we use equation (\ref{pfo}) to calculate the rear-side temperature history. 

\begin{figure}[H]
\centering
\includegraphics[width=15cm]{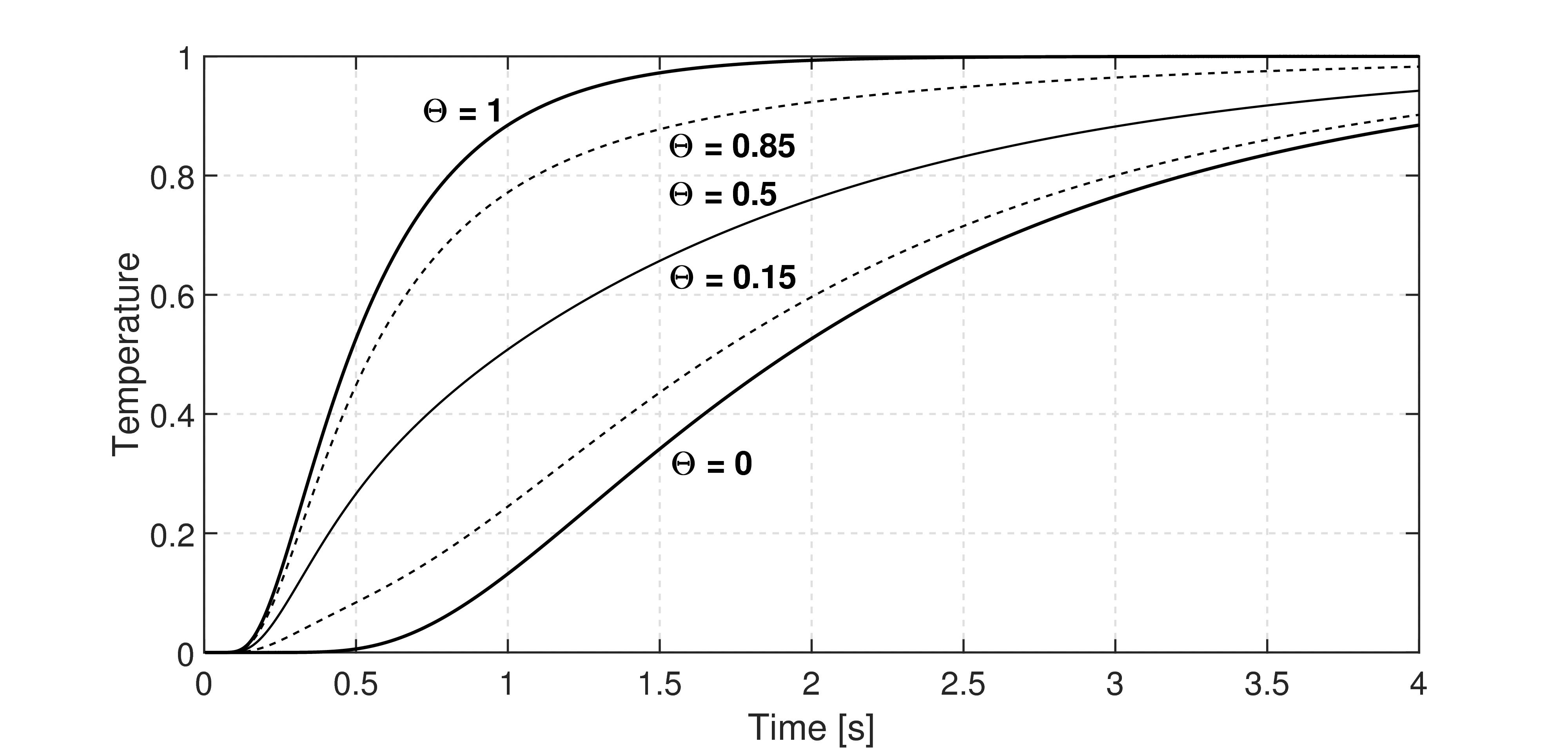}
\caption{Rear-side temperature histories.}
\label{pseudo2}
\end{figure}

So far we described two parallel heat conducting layers without direct interaction among them, however, let us suppose that they can change energy only at their rear side through a very thin layer with excellent conduction properties. Eventually, that models the role of the silver layer used in our experiments in order to close the thermocouple circuit and assure that we measure the temperature of that layer instead of any internal one from the material. Actually, the silver layer averages the rear side temperature histories of the parallel strips. 
We considered the mixing of temperature histories using the formula:
\begin{align}
p(Fo) = \Theta p_1 (a=10^{-6} \ \text{m$^2$/s}, Fo_1) + (1-\Theta) p_2 (a=2.5 \cdot 10^{-7} \ \text{m$^2$/s}, Fo_2),
\end{align}
that is, taking the convex combination of different solutions of Fourier heat equation (\ref{Fouriereq}). Fig.~\ref{pseudo2} shows a few possible cases of mixing.

\section{Outlook and summary}

This pseudo-material virtual experiment is only to demonstrate that there might be several effects causing non-Fourier behaviour of the registered temperature data. Here, the assumed mixing of ``Fourier-temperatures'' is analogous with the GK equation in sense of the hierarchy of Fourier equation: dual heat conducting channels are present and interact with each other. However, the GK equation is more general, there is no need to assume some mechanism in order to derive the constitutive equation. 

Comparing eq.~(\ref{HierGKT}) to (\ref{@17339}), the hierarchy of Fourier equation appears in a different way. While (\ref{HierGKT}) contains the zeroth and first order time derivatives of Fourier equation, the  (\ref{@17339}) instead contains its second order time and spaces derivatives. Recalling that eq.~(\ref{@17339})
\eq{@17339ag}{
\f{1}{\qcp\para^2} \9 1 { \qgamn \dot\qT - \qlam \Lapl \qT }\qddot & =
\Lapl \4 2{ \4 1 { \underbrace{ \qrho \qc - \f{ 6 \qEd \qEs \qalp^2 \qTn
}{\qEs + 2 \qEd} }_{\qgamnn} } \dot\qT - \qlam \Lapl \qT } \,.
 }
 is derived using the assumption that thermal expansion is present beside heat conduction, it becomes obvious to compare it to a ballistic (i.e., thermal expansion induced) heat conduction model. Let us consider such model from \cite{KovVan15}:
\begin{align}
\tau_1 \tau_2 \dddot T+(\tau_1 + \tau_2) \ddot T + \dot T = a \Lapl T + (\kappa^2 + a \tau_2) \Lapl \dot T, \label{BALLT}
\end{align}
where $\tau_1$ and $\tau_2$ are relaxation times. Eq.~(\ref{BALLT}) have been tested on experiments, too \cite{KovVan18}. Eventually, the GK equation is extended with a third order time derivative and the coefficients are modified by presence of $\tau_2$. On contrary to eq.~(\ref{@17339ag}), it does not contain any fourth order derivative. Actually, the existing hierarchy of Fourier equation is extended, instead of $\tau$ and $\kappa^2$ the terms $(\tau_1 + \tau_2)$ and $(\kappa^2 + a \tau_2)$ appear within (\ref{BALLT}). 

Although it is still not clear exactly what leads to over-diffusive heat conduction, the presented possible interpretations and approaches can be helpful to understand the underlying mechanism. It is not the first time to experimentally measure the over-diffusive propagation but it is to consider its size dependence. 
The simplest thermo-mechanical coupling predicts size dependence of material coefficients that can be relevant in certain cases. 
All three approaches lead to a system of partial differential equations, which can be called hierarchical.
\section{Acknowledgments}
The work was supported by the Hungarian grant National Research, Development and Innovation Office – NKFIH, NKFIH K116197, K123815, K124366, K116375.




\end{document}